\documentclass[epj,final]{svjour}
\usepackage{epsfig,float}
\restylefloat{figure}
\begin{document}
\title{Nature of the Peierls- to Mott-insulator transition in 1D}
\author{H. Fehske\inst{1} \and A.~P. Kampf\inst{2}, M. Sekania \inst{2}
\and G.~Wellein \inst{3}}
\institute{
$^1$ Institut f\"ur Physik, Ernst-Moritz-Arndt
Universit\"at Greifswald, D-17487 Greifswald, Germany\\
$^2$ Institut f\"ur Physik, Theoretische Physik III,
Universit\"at Augsburg,
86135 Augsburg, Germany\\
$^3$ RRZE, Universit\"a{}t Erlangen, 91058 Erlangen, Germany \\
}
\date{\today}
\abstract{
In order to clarify the physics of the crossover from a Peierls band insulator
to a correlated Mott-Hubbard insulator, we analyze ground-state and spectral
properties of the one-dimensional half-filled Holstein-Hubbard model using
quasi-exact numerical techniques. In the adiabatic limit the transition is
connected to the band to Mott insulator transition of the ionic Hubbard model.
Depending on the strengths of the electron-phonon coupling and the Hubbard
interaction the transition is either first order or evolves continuously
across an intermediate phase with finite spin, charge, and optical
excitation gaps.
}
\maketitle
\section{Introduction}
In quasi-one-dimensional materials like halogen-bridged transition metal
chain complexes, conjugated polymers, organic charge transfer salts, or
inorganic blue bronzes the itinerancy
of the electrons strongly competes with electron-electron and
electron-phonon  interactions, which tend to localize the charge carriers
by establishing commensurate spin- (SDW) or charge-density-wave (CDW)
ground states (GSs). At half-filling, Peierls (PI) or Mott (MI)
insulating phases are favored over the metallic state.
Quantum phase transitions between the insulating phases are
possible and the character of the electronic excitation spectra
reflects the properties of the different insulating GSs.
A controversial issue is the nature of the
PI-MI transition and whether or not only one quantum critical point
separates the PI and MI phases in purely electronic model                    
Hamiltonians~\cite{FGN99,GST00,Pati,Martin,BJK01}. Phonon dynamical
effects, which are known to
be particularly important in low-dimensional
materials~\cite{JZW99,FHW00} may further modify the transition.

In this paper we study the PI-MI quantum phase transition in the 
Holstein-Hubbard model (HHM) at half-filling. Exact numerical
methods~\cite{BWF98} are used to diagonalize the HHM on finite chains,
preserving the full dynamics of the phonons, and the density matrix
renormalization group (DMRG) technique~\cite{White}
is applied to the adiabatic
HHM  and the ionic Hubbard model.
On finite periodic chains we identify one critical PI-MI transition point in
the HHM where the site-parity of the GS changes and the excitation gap in the
optical conductivity closes. In the adiabatic limit two scenarios emerge with
a discontinuous transition at strong coupling and two subsequent continuous
transitions in the weak coupling regime with the
possibility for an intermediate
insulating phase with finite spin, charge, and optical excitation gaps.
\section{Theoretical models}
The paradigm for correlated electron-phonon systems has usually been the
one-dimensional HHM  defined by
\begin{eqnarray}\label{phm}
&&H=H_{t-U}-g\omega_0\sum_{i,\sigma}(b_i^{\dagger}+b_i^{})n_{i\sigma}+\omega_0
\sum_{i} b_i^{\dagger} b_i^{}\,\, ,\\
&&H_{t-U}=- t\sum_{i,\sigma}(c^{\dagger}_{i\sigma}c^{}_{i+1 \sigma}+\mbox{H.c.})+U\sum_i n_{i\uparrow}n_{i\downarrow}\,\, .
\label{hm}
\end{eqnarray}
$H_{t-U}$ constitutes the conventional Hubbard Hamiltonian with hopping
amplitude $t$ and
on-site Coulomb repulsion strength $U$; $c^{\dagger}_{i\sigma}$
creates a spin-$\sigma$ electron at Wannier site~$i$ and
$n_{i\sigma}=c^{\dagger}_{i\sigma}c^{}_{i\sigma}$. In~(\ref{phm}),
the second term couples the electrons locally to a
phonon created by $b_i^{\dagger}$.
Here $g=\sqrt{\varepsilon_p/\omega_0}$ is a dimensionless electron-phonon 
coupling constant, where $\varepsilon_p$ and $\omega_0$ denote the
polaron binding energy and the frequency of the optical phonon mode,
respectively.

The GS of the Holstein model  for $U=0$ is a Peierls distorted state with
staggered charge order in the adiabatic limit $\omega_0\to 0$ for any finite
$\varepsilon_p$. As in the Holstein model
of spinless fermions~\cite{BMH98,WEI98}, quantum phonon
fluctuations destroy the Peierls state for small electron-phonon interaction
strength~\cite{JZW99} -- an issue which has remained unresolved in early 
studies of the Holstein model using Monte Carlo techniques~\cite{Hirsch}. 
Above a critical threshold $g_c(\omega_0)$, the Holstein model 
describes a PI with equal spin and charge excitation gaps -- the 
characteristic feature of a band insulator (BI). 

The adiabatic limit of the HHM takes the form 
\begin{equation}
H= H_{t-U}- \sum_{i,\sigma} \Delta_{i}n_{i\sigma}+{K\over 2}\sum_{i}
\Delta_{i}^{2}
\label{hhmll}
\end{equation}
(termed AHHM); it includes the elastic energy of a harmonic lattice
with a ``stiffness constant'' $K$. In this frozen phonon approach,
$\Delta_i=(-1)^i\Delta$ is a measure of the static, staggered density
modulations of the PI phase. Eq.~(\ref{hhmll}) with $K=0$
and fixed $\Delta$ is known as the ionic Hubbard model (IHM) for which an
insulator-insulator transition was already established before, although
with controversial results regarding the possibility of an additional
intervening phase~\cite{FGN99,GST00,Pati}. Interestingly, the IHM was 
motivated originally
in quite different contexts, i.e. for the description of the
neutral to ionic transition in charge transfer salts~\cite{Nagaosa} and
ferro-electricity in transition
metal oxides~\cite{Egami,Resta}.

\section{Numerical results}
\subsection{Charge- and spin-structure factors}
\begin{figure}[t!]
\psfig{file=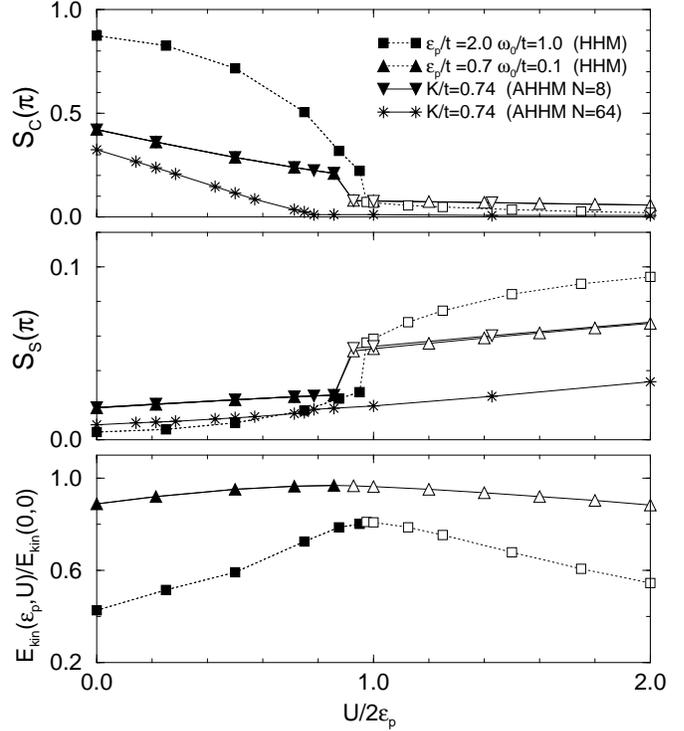,width=\linewidth}
\vspace*{3mm}
\caption{Staggered charge- (upper) and spin-structure factors (middle panel)
vs. the rescaled Hubbard interaction $U/2\varepsilon_p$. The lowest panel
displays the $U$-dependence of the kinetic energy $E_{kin}$.
Lanczos results for the HHM on an 8-site ring are given in the
adiabatic (triangles) and non-adiabatic (squares) regimes.
Lanczos ($N=8$ ring, down-triangle) and DMRG (open 64-site chain, stars) 
results are shown for the AHHM with $K=0.74$. Open (closed) symbols belong 
to GSs with site-parity $P=-1$ (+1).}
\label{fig:fig1}
\end{figure}

In order to establish the GS properties of the above models and the existence
of the PI-MI transition we start with the evaluation of the staggered
charge- and spin-structure factors $S_c(\pi)$ and $S_s(\pi)$, respectively,
\begin{eqnarray}
S_{c}(\pi) &=& \frac{1}{N}\sum_{j,\sigma\sigma'} (-1)^j\langle
(n_{i\sigma}-\frac{1}{2} )(n_{i+j,\sigma'}-\frac{1}{2})\rangle\, , \nonumber\\
S_{s}(\pi) &=& \frac{1}{N}\sum_{j} (-1)^j\langle S_i^zS_{i+j}^z\rangle\,\, ,
\, S_i^z={1\over 2}(n_{i\uparrow}-n_{i\downarrow})\, .\nonumber
\label{scs}
\end{eqnarray}
Results for the $U$-dependences of $S_c(\pi)$ and $S_s(\pi)$ on an 8-site HHM
ring are shown in Fig.~\ref{fig:fig1} for two different phonon frequencies,
corresponding to adiabatic and non-adiabatic regimes.  The PI regime is
characterized by a large (small) charge- (spin-) structure factor.
Also shown in Fig.~\ref{fig:fig1} are results for the AHHM [$N=8$ (Lanczos), 
$N=64$ (DMRG)] which will be discussed in Sec. 3.3
Increasing $U$ at fixed $\varepsilon_p$ and $\omega_0$, Peierls CDW order is
suppressed as becomes manifest from the rapid drop of $S_c(\pi)$ which decreases
nearly linearly in the adiabatic regime, but
its initial decrease is significantly weaker for higher phonon frequencies.
The disappearance of the charge ordering signal is
accompanied by a steep rise in $S_s(\pi)$
indicating enhanced antiferromagnetic correlations in the
MI phase. The data for $S_c(\pi)$ provide evidence for
a critical point $U_c$ at which the  CDW order
disappears: rather abruptly for adiabatic and smoothly for
non-adiabatic phonon frequencies. Above $U_c$ the low-energy
physics of the system is qualitatively similar to the pure Hubbard chain; it
is governed by gapless spin and massive charge excitations. As compared
to the PI phase, the local magnetic moment
$L_i(U/t, \varepsilon_p/t, \omega_0/t)\propto\langle (S_i^z)^2 \rangle$
is strongly enhanced (e.g.  $L_i(8,2,1)/L_i(0,2,1)\simeq 16.4 $).

\begin{figure}[h!]
\psfig{file=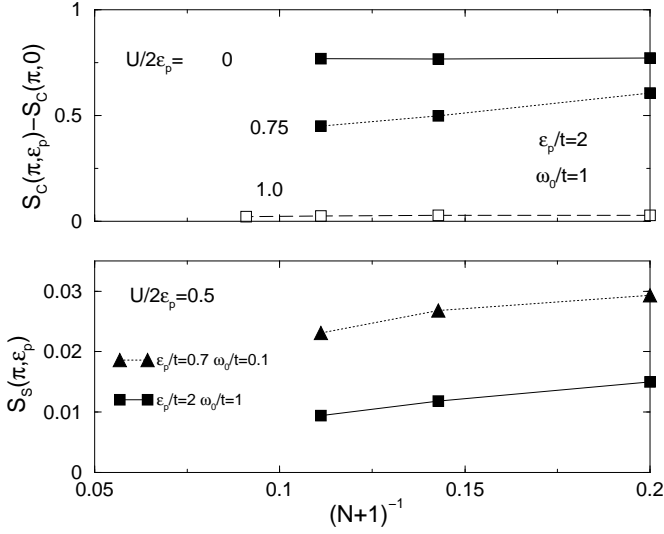,width=\linewidth}
\vspace*{3mm}
\caption{Finite-size scaling of charge- (upper panel) and
spin- (lower panel) structure factors in the HHM. }
\label{fig:fig2gw}
\end{figure}

We emphasize the weak finite-size dependence of the exact
diagonalization data for $S_c(\pi)$ in both the
strong-CDW Peierls and MI phases (see the upper panel
of Fig.~\ref{fig:fig2gw}). The variation of $S_s(\pi)$
with the system size $N$ points to a vanishing
spin-structure factor in the CDW state, i.e. for $U < U_c$
but, of course, it is beyond our current numerical capabilities to perform
a real finite-size analysis for the HHM with dynamical phonons.
In the adiabatic limit, we expect a finite $S_s(\pi)$
in the MI phase as for the corresponding SDW
state of the so-called extended Hubbard model with $U>2V$~\cite{FS84,H84}.

\begin{figure}[b!]
\psfig{file=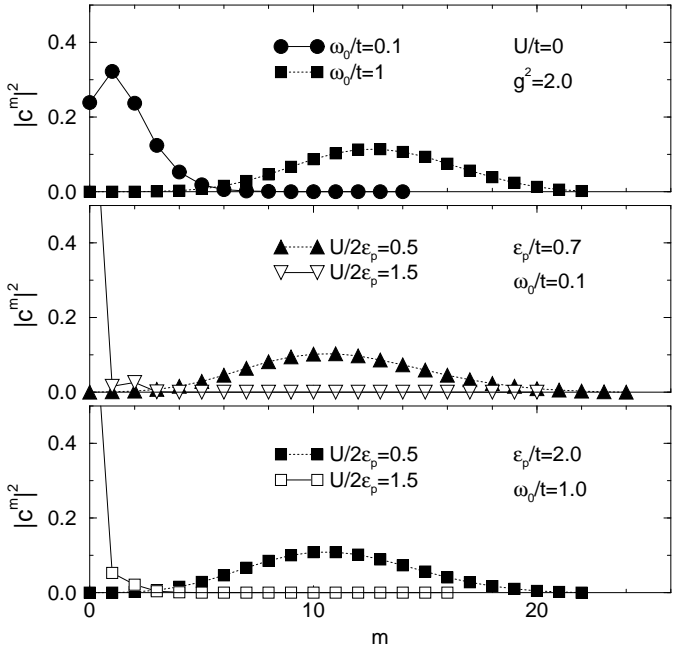,width=\linewidth}
\vspace*{3mm}
\caption{Phonon distribution in the GS of the HHM for various model
parameters. In the MI state (open symbols) the weight of the
zero-phonon state is almost one, \mbox{$|c^m|^2\simeq 1$}. }
\label{fig:fig3gw}
\end{figure}

To discuss the lattice dynamical effects in some more detail
we show in Fig.~\ref{fig:fig3gw} the weights of $m$-phonon states
in the GS of the HHM at different interaction strengths.
First of all  Fig.~\ref{fig:fig3gw} demonstrates that
our phonon Hilbert space truncation procedure is well-controlled
in the sense that states with larger numbers of phonons, as accounted
for in the calculations, have negligible spectral weight.
Of course, the number of phonons
which have to be taken into account strongly depends on the
physical situation. Whereas the GS of the MI is almost a
zero-phonon state, multi-phonon states become increasingly
important if $U$ is reduced ($\varepsilon_p$ is enhanced)
in the PI regime.

For small  $U$ and low phonon frequencies the PI phase appears for
$g>g_c(\omega_0)$, even if the ratio $\lambda=\varepsilon_p/2t$ is small.
In the GS of such a conventional BI phonons populate
predominantly the $q=0$ and 
$\pi$ modes, but the total number of phonons involved in the creation of
the  Peierls-distorted  CDW is rather small. On the contrary, 
for large phonon frequencies, $g>g_c(\omega_0)$ implies 
$\lambda \gg 1$ and we observe a multi-phonon GS (cf.
Fig.~\ref{fig:fig3gw}). As a consequence the electrons are heavily dressed 
by phonons, forming bipolarons in real space, which lower their 
energy by ordering in a staggered CDW. Therefore, in the non-adiabatic 
strong electron-phonon coupling regime, the system is classified rather as a 
charge-ordered bipolaronic insulator than as a BI.
The kinetic energy is much more suppressed for the bipolaronic
than for the band PI (cf. lower panel of Fig.~\ref{fig:fig1}). Since
SDW  correlations reduce $E_{kin}$ as well, the kinetic energy reaches a
maximum when the system crosses from the PI to the MI regime.

\begin{figure}[b!]
\vspace{0mm}
\psfig{file=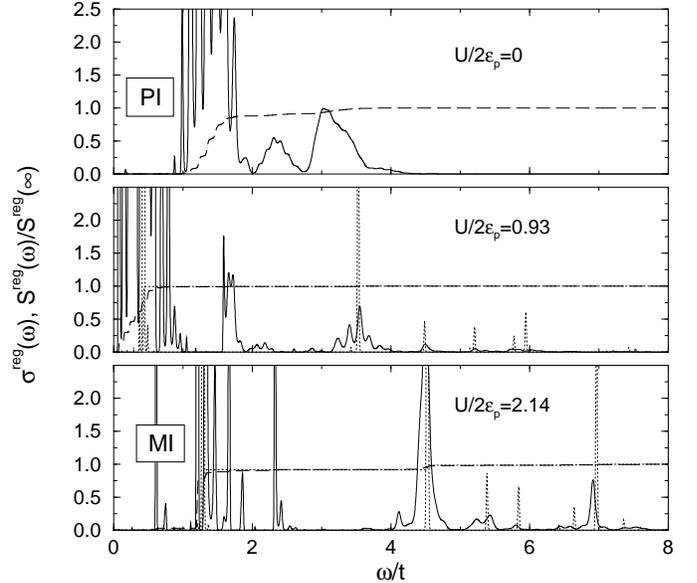,width=\linewidth}
\caption{Optical conductivity in the 8-site HHM for \mbox{$\omega_0\!=\!0.1t$}
and $g^2\!=\!7$.
Top panel: PI phase for $U=0$; middle panel: near criticality
$U\sim U_{opt}$; lower panel: MI phase for $U=3t$. Dashed lines give
the normalized integrated spectral weights $S^{reg}(\omega)$.
The lower two panels include $\sigma^{reg}$ for $g=0$
(dotted lines), i.e. for the pure Hubbard chain.}
\label{fig:fig2}
\end{figure}

\subsection{Optical response}

Valuable insight into the nature of the PI-MI transition is obtained from
symmetry considerations~\cite{GST00,BJK01}. The BI-MI transition of the IHM
on finite lattices was shown to be connected to a GS level crossing with a
site-parity change, where the site inversion symmetry operator $P$ is defined
by $Pc_{i\sigma}^\dagger P^\dagger=c_{N-i\sigma}^\dagger$ with $N=4n$ for
$i=0,...,N-1$. This feature will become evident in the regular part of
the optical conductivity at $T=0$,
\begin{equation}
\sigma^{reg}(\omega)\!=\!\frac{\pi}{N}\!\sum_{m\neq 0}\!
\frac{|\langle\psi_0|\hat{j}
|\psi_m\rangle|^2}{E_m-E_0}\,\delta(\omega\!-\!E_m\!+\!E_0).
\label{sigmareg}
\end{equation}
Here, $|\psi_0\rangle$ and $|\psi_m\rangle$ denote the GS and excited states,
respectively, and $E_m$ the corresponding eigenenergies. Importantly, the
current operator $\hat{j}=-{\rm i}e t\sum_{i\sigma}(c_{i\sigma}^{\dagger}
c_{i+1\,\sigma}^{}-c_{i+1\,\sigma}^{\dagger}c_{i\sigma}^{})$ has finite
matrix elements between states of different site-parity only.

The evolution of the frequency dependence of $\sigma^{reg}(\omega)$
from the PI to the MI phase with increasing $U$ is illustrated in
Fig.~\ref{fig:fig2}. In the PI regime the electronic excitations are gapped
due to the pronounced CDW correlations. The broad optical absorption band 
for $U=0$ results from particle-hole excitations across the BI gap which 
are accompanied by multi-phonon absorption and emission processes. The shape 
of the absorption band reflects the phonon distribution function in the GS. 
Excitonic gap states may occur in the process of structural relaxation. At
$U_{opt}$ the optical gap $\Delta_{opt}$ closes, and due to 
the selection rules for optical 
transitions this necessarily implies a GS level crossing with a site-parity 
change. We have explicitly verified that the GS site parity in the PI phase is 
$P=+1$ and $P=-1$ in the MI phase (see also Fig.~\ref{fig:fig1}). 
For the HHM on finite rings $U_{opt}$ is identical 
to the critical point where $S_c(\pi)$ sharply drops.

For the adiabatic phonon frequency used in Fig.~\ref{fig:fig2} the phonon 
absorption threshold is small and, because the GS is a multi-phonon state,
we find a gradual linear rise of the integrated spectral weight 
$S^{reg}(\omega)=\int_0^\omega \sigma^{reg}(\omega')\,{\rm d}\omega'$.
$S^{reg}(\omega)/S^{reg}(\infty)$ is a natural measure for  
the relative weight of the different optical absorption processes. 
In contrast, in the non-adiabatic regime
($\omega_0\geq t$), the lowest optical excitations have mainly pure 
electronic character in the vicinity of $U_{opt}$.
As a result the gap is closed by a state having large electronic spectral 
weight. 

In the MI phase the optical gap is by its nature a correlation gap. The lower 
panel in Fig.~\ref{fig:fig2}  shows clearly that $\sigma(\omega)$ of the HHM
in the MI phase is dominated by excitations which can be related to those of
the pure Hubbard model. In addition, phononic sidebands with low spectral
weight and phonon-induced gap states appear.

\subsection{Phase diagram in the adiabatic limit}
The above results for the HHM establish the PI-MI phase transition scenario on
small rings and trace it to the level crossing of the two site-parity
sectors. In order to draw conclusions about the phase diagram in the adiabatic
regime we exploit the connection to the AHHM.

\begin{figure}[t!]
\vspace{0mm}
\psfig{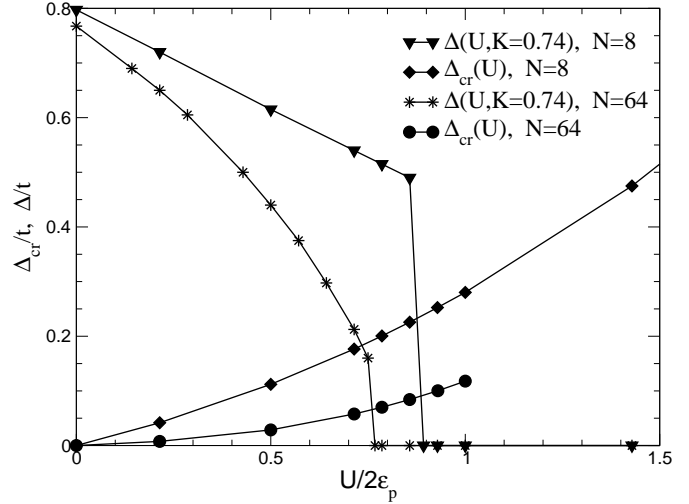}
\caption{Level crossing line $\Delta_{cr}(U)$ of the
IHM for an 8-site ring (diamonds) and from extrapolating Lanczos data for
$N\leq 14$ to a 64-site chain (circles). In addition: ionic potential
strength $\Delta(U,K)$ of the AHHM for an 8-site ring
(triangles) and on an open 64-site chain (stars, DMRG results) for $K=0.74$.}
\label{fig:fig3}
\end{figure}

The magnitude of $S_c(\pi)$ in the HHM for $U=0$ and $\omega_0=0.1t$ allows a
straightforward way to fix the stiffness constant $K$ in Eq.~(\ref{hhmll}).
Using the result of the AHHM for $S_c(\pi)$ at $U=K=0$ we determine first the
ionic potential strength $\Delta_0$ by the requirement that
$S_c^{IHM}(\pi,\Delta_0)=S_c^{HHM}(\pi)$ for the same chain length and 
periodic boundary conditions. In a second step, the GS energy of the AHHM, 
$E_0(K,\Delta,U=0)$, determines $K$ by the criterion that $E_0$ is minimized 
for $\Delta=\Delta_0$. We thereby obtain $K=0.74$, which is henceforth kept 
fixed when the interaction
$U$ is turned on. For each value of $U$, the ionic potential strength of the
AHHM is then obtained by minimizing $E_0(K,\Delta,U)$ with
respect to $\Delta$, yielding $\Delta=\Delta(U,K)$ as shown in
Fig.~\ref{fig:fig3} (triangles). The resulting structure factors for the AHHM 
are plotted in Fig.~\ref{fig:fig1}, too, and agree very accurately with the 
8-site HHM ring data for $\omega_0/t=0.1$. 
This agreement reconfirms numerically that the 
AHHM is indeed the appropriate effective model to describe the CDW phase of 
the HHM in the adiabatic limit.
The drop in $S_c(\pi)$ at the transition point results from a
discontinuous vanishing of $\Delta(U,K)$ (see Fig.~\ref{fig:fig3}).
The large charge structure factor $S_c(\pi)$ below $U_c$ and the enhancement 
of the spin structure factor $S_s(\pi)$ above $U_c$
as well as the sharp changes at the transition point find a natural 
explanation with the results for $\Delta(U,K)$ in Fig.~\ref{fig:fig3}. 
Below the transition $\Delta$ is finite implying long range CDW order in the 
GS. At the transition point $\Delta$ vanishes
discontinuously and thereby the AHHM reduces to the pure
Hubbard model ($\Delta=0$).

Given the value for the stiffness constant $K$ we also plot in
Fig.~\ref{fig:fig3} $\Delta(U,K)$ obtained from DMRG
on an open chain of length $N=64$. For comparison, the corresponding results 
for $S_c(\pi)$ and $S_s(\pi)$ in the AHHM are shown in Fig.~\ref{fig:fig1}, 
too (stars). $S_c(\pi)$ decreases smoothly and almost linearly; although 
unresolved on
the vertical scale in Fig.~\ref{fig:fig1} the transition remains discontinuous
as a consequence of the results for $\Delta(U,K)$ in Fig.~\ref{fig:fig3}. In 
contrast to the behavior of the 8-site chain, $\Delta(U,K)$ here decreases 
more smoothly with increasing $U$ and
vanishes discontinuously near $U/2\varepsilon_p\approx 0.75$.
The small discontinuous increase in $S_s(\pi)$ at the transition is also 
hardly resolved for the 64-site chain in contrast to the 8-site chain data.
The discontinuous nature of the PI-MI transition in the AHHM for
$\Delta_i=(-1)^i\Delta$ is obvious in the atomic limit $t=0$ where
$\Delta=1/K$ for $U<U_c=1/K$ and $\Delta=0$ for $U>U_c$. As verified above, 
the first order nature persists for finite small $t$, i.e. in the strong 
coupling regime $U$, $K^{-1}\gg t$.

Also shown in Fig.~\ref{fig:fig3} is the level crossing line $\Delta_{cr}(U)$
of the IHM for $N=8$ (diamonds) and $N=64$ (circles) chain. $\Delta_{cr}(U)$
for  $N=64$ was obtained from extrapolating Lanczos results for rings of up to
14 sites to a 64-site chain~\cite{boundary}. Importantly, $\Delta(U,K)$ and
$\Delta_{cr}(U)$ do not intercept because $\Delta(U,K)$ jumps to zero before
reaching the level crossing point of the IHM.

\begin{figure}[t!]
\vspace{0mm}
\psfig{file=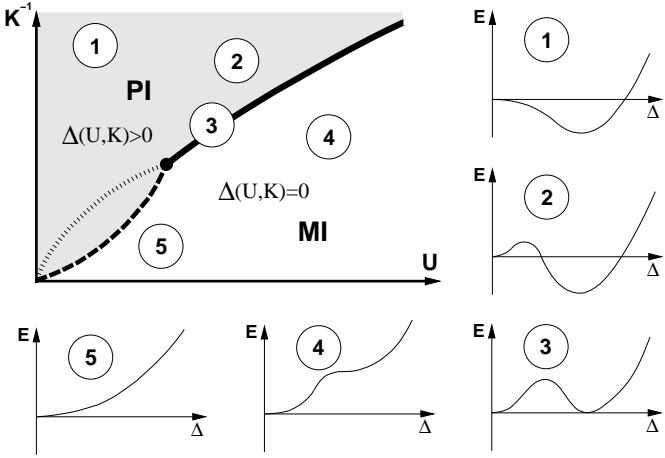 ,width=\linewidth}
\caption{Insets (1) - (5): Evolution of the ground-state energy vs. $\Delta$ 
in the AHHM in different regions of the $(K^{-1},U)$ parameter plane. From 
the variations in $E(\Delta)$ a crossover from a discontinuous PI (with
$\Delta>0$) to MI ($\Delta=0$) transition to a second order transition is 
deduced. Main figure: Phase diagram of the AHHM; the solid line represents a 
discontinuous, first order and the dashed line a continuous second order 
transition. These results summarize Lanczos data for a 14-site AHHM chain 
with periodic or open boundary conditions. Detailed runs where performed for 
$U=0.3t$ and $U=5t$. A possible additional continuous transition (dotted line) 
between two insulating phases with finite $\Delta$ is indicated as well.}
\label{fig:fig6}
\end{figure}

The DMRG results presented in Fig.~\ref{fig:fig3} for $N=64$ raise the question
whether the discontinuous transition in the AHHM can turn into a continuous
transition on approaching the weak coupling regime by increasing the stiffness
constant $K$. Indeed, as we have explicitely verified by exact diagonalization
of a periodic (and open too) AHHM ring of length $N=14$, the transition is 
second order in the regime $U$, $K^{-1}\ll t$. The corresponding Lanczos 
results for the variation of the GS energy vs. $\Delta$ in the 
$(K,U)$-parameter plane are summarized in Fig.~\ref{fig:fig6}. Detailed 
$K$-scans were performed for weak ($U=0.3t$) and strong ($U=5t$) Hubbard 
interaction. The evolution of $E(\Delta)$ in the AHHM in fact reveals that 
the transition from the PI to the MI phase (sequence $(2)-(3)-(4)$) occurs 
discontinuously at strong coupling $K^{-1},U\gg t$, while the transition 
follows a Ginzburg-Landau-type behavior for a second order phase transition at 
weak coupling (sequence $(1)-(5)$ in Fig.~\ref{fig:fig6}).

Due to the continuous decrease of $\Delta(U,K)$ at weak coupling $\Delta(U,K)$
necessarily intercepts the $\Delta_{cr}(U)$ line of the IHM~\cite{boundary}.
This intercept marks the point $U_{opt}$ when the site-parity sectors become
degenerate and the optical absorption gap $\Delta_{opt}$ disappears. This
situation therefore implies the existence of an intermediate region
$U_{opt}<U<U_s$ with finite $\Delta$, where $U_s$ marks the point where
$\Delta$ continuously vanishes. Since $\Delta=0$ for $U>U_S$, i.e. when the
AHHM reduces to the Hubbard model, the spin gap vanishes at $U_s$. The
intermediate insulating phase thus has finite spin, charge, and optical
excitation gaps. For weak coupling the PI-MI transition therefore evolves
across two critical points $U_{opt}$ and $U_s$. The $U$ vs. $K^{-1}$ phase
diagram contains a multicritical point at which a first order line splits into
two continuous transition lines. The additional transition line is also 
indicated in Fig.~\ref{fig:fig6} (dotted line). For weak $U$ at fixed $K$ the 
transition at $U_{opt}$ is expected to be of Kosterlitz-Thouless type since 
it corresponds to the merging of the energies of the two site-parity sectors; 
the CDW vanishes in a
second order type transition at $U=U_s$. The $E(\Delta)$ behavior, however, can
only detect the boundary to the MI phase of the AHHM where the GS energy is
minimized for vanishing $\Delta$.

\begin{figure}[t!]
\vspace{0mm}
\psfig{file=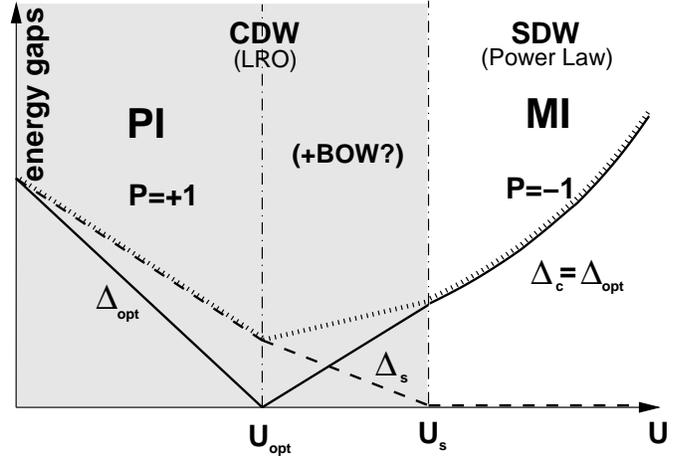 ,width=\linewidth}
\caption{Qualitative behavior of the excitation gaps versus $U$ in the weak 
coupling regime of the AHHM $U, K^{-1}\ll t$. Solid line: optical excitation 
gap $\Delta_{opt}$, dotted line: charge gap $\Delta_c$, dashed line: spin gap
$\Delta_s$. PI phase: $\Delta_c =\Delta_s$ and site parity $P=+1$; MI phase: 
$\Delta_{opt}=\Delta_c$ and $P=-1$.}
\label{fig:fig4}
\end{figure}

We summarize these findings in the diagram for the excitation gaps shown in
Fig. \ref{fig:fig4},
\begin{eqnarray}\label{gapc}
\Delta_c &=& E_0(N/2+1,N/2)+E_0(N/2-1,N/2)\nonumber\\
 &&-2E_0(N/2,N/2)\,,\\
\Delta_s &=& E_0(N/2+1,N/2-1)-E_0(N/2,N/2)\,,
\label{gaps}
\end{eqnarray}
where $E_0(N_\uparrow,N_\downarrow)$ is the GS energy of the system with
$N_\uparrow$ spin-up and $N_\downarrow$ spin-down electrons.
In the Peierls BI phase for $U<U_{opt}$ the spin and
charge gaps, are equal and finite and remarkably $\Delta_{opt}\neq\Delta_c$
(for a similar conclusion in the IHM see \cite{Qin}). At $U=U_{opt}$ the
site-parity sectors become degenerate, $\Delta_{opt}=0$ but remarkably
$\Delta_c=\Delta_s>0$. For $U\geq U_s$ the usual MI phase of the half-filled 
Hubbard chain with $\Delta_{opt}=\Delta_c>\Delta_s=0$ is realized. For strong 
coupling, when the PI to MI transition is first order, $U_{opt}=U_s$, the spin
gap discontinuously disappears at the transition and the optical gap jumps 
from zero to the finite charge gap value of the Hubbard chain. In
weak coupling there exists an intermediate region $U_{opt}<U<U_s$ in which all
excitation gaps are finite. The CDW persists for all $U<U_s$. The site-parity
eigenvalue is $P=+1$ in the PI and $P=-1$ in the MI phase.

The insulating, intermediate phase at weak coupling as identified above
remains yet to be characterized. For the insulator-insulator phase
transition(s) in the IHM Fabrizio et al. proposed the existence of an
intermediate phase with a long range bond order wave (BOW) based on a
bosonization analysis~\cite{FGN99}. BOW order is characterized by a finite
expectation value of the staggered bond charge $B={1\over N}\sum_{i\sigma}
(-1)^i\langle c^\dagger_{i\sigma}c^{{\phantom{\dagger}}}_{i+1\,\sigma}+H.c.
\rangle$. Some positive numerical evidence has indeed been reported for
enhanced BOW correlations above the level crossing transition in the
IHM~\cite{Pati,Martin,BJK01,Torio01}. Yet, these results have remained
ambiguous so far and no consensus has been reached about the existence of long
range BOW order in the IHM. In the attempt to search for BOW order the DMRG
calculations, which for numerical accuracy reasons are predominantly performed
on open chains, suffer from the fact that Friedel-like bond charge density
oscillations are induced by the chain ends already for the pure Hubbard
chain~\cite{BJK01}. The identification of BOW order in the IHM or AHHM by DMRG
on open chains therefore requires a delicate subtraction procedure to
discriminate a BOW signal from the edge induced bond charge oscillations of
the Hubbard chain. We have nevertheless attempted to search for BOW
correlations in the weak coupling regime of the AHHM, where the continuous
nature of the transition into the MI phase was established by the Lanczos
results on the periodic or open 14-sites chain, i.e. these calculations 
naturally focused on the weak-$U$ regime ($U<t$). This weak coupling regime is
notoriously hard for numerical evaluations; unfortunately the numerical
accuracy needed to allow a firm conclusion about the presence or absence of a
BOW signal could not be achieved within our DMRG runs.

While a confirmation is thus still lacking BOW order remains a vivid candidate
order in coexistence with a CDW to characterize the intermediate phase in the
AHHM at weak coupling. We furthermore note that if the existence of a BOW is
verified in the AHHM, its phase diagram would be
remarkably similar to the extended Hubbard model with nearest neighbor Coulomb
repulsion with an intervening BOW phase in the crossover between the CDW and
MI phases at weak coupling \cite{Na,Ca}.

\section{Conclusions}
In summary, we have found a PI-MI transition in the HHM above a threshold
electron-phonon coupling.
The transition results from a GS level crossing with a change in the
GS site-parity eigenvalue. In the adiabatic limit two scenarios emerge with a 
discontinuous PI-MI transition for $U,K^{-1}\gg t$, and two continuous 
transitions for weak coupling $U,K^{-1}\ll t$ with an intermediate phase where 
CDW order persists. In the non-adiabatic regime our structure factor data 
indicate that the PI-MI transition proceeds continuously.

\section*{Acknowledgments}
We appreciate discussions with A.R. Bishop, F. G\"ohmann, G.I. Japaridze, and A.
Wei{\ss}e. Calculations were performed at the NIC J\"ulich and the LRZ 
M\"unchen, supported by the Bavarian Network for High-Performance Computing 
(KONWIHR). H.F. and A.P.K. acknowledge support through SPP 1073 and SFB 484 of 
the Deutsche Forschungsgemeinschaft, respectively.


\begin{thebibliography}{99}
\bibitem{boundary} To obtain $\Delta_{cr}(U)$ for rings 
with $N=4m+2$ and $N=4m$ sites we used antiperiodic and periodic 
boundary conditions, respectively.
\end{thebibliography}
\end{document}